\documentclass[twocolumn,showpacs,preprintnumbers,amsmath,amssymb]{revtex4}
\usepackage{graphicx}% Include figure files
\usepackage{dcolumn}% Align table columns on decimal point
\usepackage{bm}% bold math

\begin{document}

\title{
Localized tendency for the superfluid and Mott insulator state\\
%Localized tendency for quantum phase transition
in the array of dissipative cavities
}

\author{
Lei Tan$^{1}$, Ke Liu$^{1}$, Chun-Hai Lv$^{1}$, W. M. Liu$^{2}$ }

\affiliation{
$^{1}$Institute of Theoretical Physics, Lanzhou University, Lanzhou 730000, China\\
$^{2}$Beijing National Laboratory for Condensed Matter Physics, Institute of Physics, Chinese Academy of Sciences, Beijing 100190, China
}
\date{\today}

\begin{abstract}
The features of superfluid-Mott insulator phase transition in the
array of dissipative cavities is analyzed. Employing a kind of
quasi-boson and a mean-filed approach, we show analytically how
dissipation and decoherence influence the critical behaviors and
the time evolution of the system. We find that there is a
localized tendency, which could lead to the break of superfluidity
for a superfluid state and suppress the appearance of the
long-range order form a Mott state. Eventually, a collection of
mixture states localized on each site will arise.
\end{abstract}

\pacs{42.50.Pq, 42.60.Da, 64.70.Tg, 03.65.Yz}
% 42.50.Pq Cavity quantum electrodynamics; micromasers
% 42.60.Da Resonators, cavities, amplifiers, arrays, and rings
% 64.70.Tg Quantum phase transitions
% 03.65.Yz Decoherence; open systems; quantum statistical methods

\maketitle

One of the remarkable applications of coupled cavity arrays
is to realize so called quantum simulators.
Due to the controllability of atomic and optical systems,
it could be useful to attack some unclear physics and
to explore new phenomenon in quantum many-body systems~[\onlinecite{SC1, SC2, SC3, CCA}].
In particular, over the past years the experimental progresses in
controlling optical system~[\onlinecite{C_M, C_C, C_S, C_B}]
and in fabricating large-scale arrays of
high-$Q$ cavities~[\onlinecite{array1, array2}]
make this potential application close to reality.
However, the quantum optical system is typically driven by an external field
and coupled to the environment~[\onlinecite{open1, open2}],
which bring the system out
of equilibrium and profoundly affected the dynamics of interest~[\onlinecite{N-E1, N-E2}].
New important questions thus arise and need to be clarified, such as
whether the link between the initial ideas of cavity arrays as
quantum simulators and the realistically experimental conditions
is hold, and how dissipation and decoherence would behave in the system.

In this paper, we propose answers to these questions by investigating
the superfluid(SF)-Mott insulator phase transition in the array of
dissipative cavities.
Provided the external time dependence is much slower than
the internal frequencies of system,
we show that there are still two fundamentally different
quantum states, i.e.
photons localized in each cavity(Mott-like)
and delocalized cross the lattices(SF-like).
However, a localized tendency does exist for both of these two states.
Specifically,
the ratio of the the intercavity coupling and on-site interaction strength
for the transition from Mott-like region
to SF-like region is modified by a time dependent increment.
Moreover, the superfluidity of a initial SF state will break
owing to the decay of the off-diagonal long-range order.

Consider a system consisted of atoms and cavities
coupled weakly to a bosonic environment at zero
temperature.
As the dimension of individual cavities is generally much
smaller than their spacing,
we assume the photons emitted from each cavity
are uncorrelated. The total Hamiltonian therefore reads
\begin{equation}
H=H_{s}+H_{bath}+H_{coup}.
\end{equation}
where $H_{s}$ is the Hamiltonian for the system,
$H_{bath}=\sum_{j}\sum_{\alpha,k}\omega_{k_{\alpha}}r^{\dag}_{j,k_{\alpha}}r_{j,k_{\alpha}}$
the Hamiltonian for environment, and
$H_{coup}=\sum_{j}\sum_{\alpha,k}(\eta^{\ast}_{k_{\alpha}}r^{\dag}_{j,k_{\alpha}}\alpha_{j}+h.c.)$
the coupled term.
$\alpha=a, c$ labels the operators and physical quantities associated with atoms and
cavities, respectively. $\omega_{k_{\alpha}}$ denotes the frequency of
environmental modes,
$r^{\dag}_{j,k_{\alpha}}$ and $r_{j,k_{\alpha}}$the creation and annihilation operators of
quanta in the $k_{\alpha}$th model
at site $j$, and $\eta_{k_{\alpha}}$ the coupling strength.
Here we set $\hbar=1$.

\begin{figure}[b]
\includegraphics[clip]{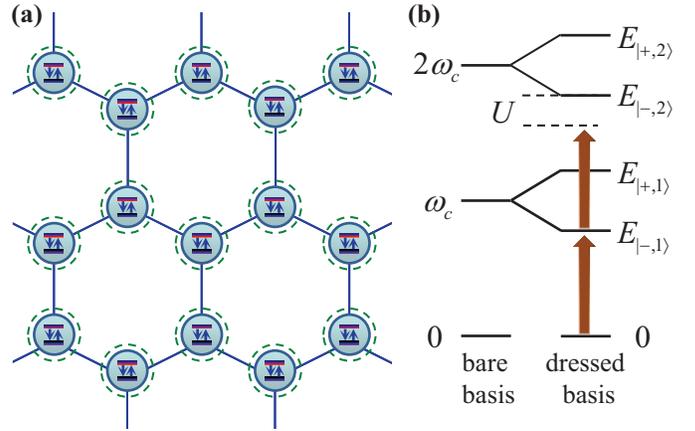}
\caption{
\label{fig:epsart}
A type of possible topologies for two-dimensional cavity arrays.
(a) Individual cavities are coupled resonantly to each other due to the
overlap of the evanescent fields. Each cavity contains a two-level system
coupled strongly to the cavity field and
immerses in a bosonic bath(marked by the dash line).
(b) Energy eigenvalues of individual cavity-atom system
on each site. $\omega_{c}=\omega_{a}$ is assumed for simplify.
The anharmonicity of the Jaynes-Cummings energy levels
can effectively provide an on-site repulsion $U$
to block the absorption for the next photon.
}
\end{figure}
The system we modelled, as depicted in Fig. 1, is a
two-dimensional array of resonant optical cavities, each embedded
with a two-level (artificial)atom coupled strongly to the cavity
field. The possible realizations such as photonic bandgap cavities
and superconducting stripline resonators et
al.~[\onlinecite{CCA}]. With $\omega_{a}$ and $\omega_{c}$ being
the frequency of atom transition and cavity mode respectively, in
the rotating wave approximation(RWA), such individual atom-cavity
system on site $j$ is well described by the Jaynes-Cummings
Hamiltonian, $
H^{JC}_{j}=\omega_{a}a^{\dag}_{j}a_{j}+\omega_{c}c^{\dag}_{j}c_{j}+\beta(a^{\dag}_{j}c_{j}+h.c.)
$. Here $a^{\dag}_{j}$ and $a_{j}$($c^{\dag}_{j}$, $c_{j}$) are
atomic(photonic) rasing and lowering operators, respectively,
$\beta$ the coupled strength. In the grand canonical ensemble,
$H_{s}$ is therefore given by combing $H^{JC}_{j}$ with photonic
hopping term and chemical potential term,
\begin{equation}
H_{s}=\sum_{j}H^{JC}_{j}-\sum_{\langle j,j'\rangle}\kappa_{jj'}c^{\dag}_{j}c_{j'}-\sum_{j}\mu n_{j}.
\end{equation}
$\kappa_{jj'}$ is the photonic hopping rate between cavities.
Since the evanescent coupling between cavities decreases with the distance exponentially,
we restrict the summation $\sum_{\langle j,j'\rangle}$ running
over the nearest-neighbors.
$n_{j}=a^{\dag}_{j}a_{j}+c^{\dag}_{j}c_{j}$
is the total number of atomic and photonic
excitations on site $j$.
$\mu$ is the chemical potential, where the assumption $\mu=\mu_{j}$
for all sites has been made.

Due to the strongly coupling, as shown in Fig. 1(b), the resonant
frequencies of individual atom-cavity system are split into
\begin{equation}
E_{|\pm,n\rangle}=n\omega_{c}\pm\sqrt{n\beta^{2}+\frac{\Delta^{2}}{4}}-\frac{\Delta}{2},
\end{equation}
where $|\pm,n\rangle$ labels the positive(negative) branch of dressed states,
$\Delta=\omega_{c}-\omega_{a}$ is the detuning.
The anharmonicity of the Jaynes-Cummings energy levels can effectively
provide a on-site repulsion. For instance,
the resonant excitation by a photon with frequency $E_{|\pm,1\rangle}$ will
prevent the absorption of a second photon at $E_{|\pm,1\rangle}$,
which is the striking effect known as photon blockade~[\onlinecite{C_B}].
It is therefore feasible to realize a quantum simulator in terms of
the system described by Eq. (2).
This so called Jaynes-Cummings-Hubbard(JCH) mode is recently suggested
by Greentree et al.~[\onlinecite{SC2}].

However, the situation changes dramatically
once taking the coordinates of environment into consideration,
as described by Hamiltonian(1).
A non-equilibrium dynamics for open quantum many-body system do arise,
which is a formidable task to solve.
Here we propose a treatment to
eliminate those external degrees of freedom.
To approach this, we rewrite Hamiltonian(1) as
\begin{equation}
H=H_{local}-\sum_{\langle j,j'\rangle}\kappa_{jj'}c^{\dag}_{j}c_{j'}-\sum_{j}\mu n_{j},
\end{equation}
where
$H_{local}=\sum_{j}H^{JC}_{j}+H_{bath}+H_{coup}$.

First considering the case that the $j$th cavity contained a initial photon interacts with
a bath, the dynamics is governed by
\begin{equation}
H_{j}=\omega_{c}c^{\dag}_{j}c_{j}+\sum_{k}\omega_{k_{c}}r^{\dag}_{j,k_{c}}r_{j,k_{c}}
+\sum_{k}(\eta^{\ast}_{k_{c}}r^{\dag}_{j,k_{c}}c_{j}+h.c.).
\end{equation}
We denote its eigenvalue as $\omega$ and
expand the eigenvector $|\phi_{j}\rangle$ as
$
|\phi_{j}\rangle=e_{c}c^{\dag}_{j}|\emptyset\rangle
+\sum_{k}e_{k}r^{\dag}_{k_{c}}|\emptyset\rangle
$.
$e_{c}$ and $e_{k}$ are the probability amplitudes for the excitation occupied by
cavity field and environment, respectively. $|\emptyset\rangle$ denotes the vacuum state.
Deducing the equations of these two amplitudes, one can express
$e_{k}$ in terms of $e_{c}$ and, under the Born-Markov approximation,
integrate out the degrees of freedom of environment, and obtain
\begin{equation}
(\omega_{c}+\delta\omega_{c}-i\gamma_{c})e_{c}=\omega e_{c}.
\end{equation}
$\delta\omega_{c}$ is known as an analog to the Lamb shift in atomic physics
and significantly small when the coupling to environment is weak.
$\gamma_{c}$ is the decay rate and indicates a finite lifetime of cavity mode~[\onlinecite{book}].

This motivates us to introduce a quasi-boson described by $C_{j}$ with
a complex eigenfrequency
$
\Omega_{c}=\omega_{c}-i\gamma_{c}
$,
where $\delta\omega_{c}$ has been absorbed into $\omega_{c}$,
to effectively describe the open system described by Hamiltonian (5) in terms of
\begin{equation}
H^{eff}_{j}|\phi_{j}\rangle=\omega^{eff}_{j}|\phi_{j}\rangle,
\end{equation}
with the effective Hamiltonian
$
H^{eff}_{j}=\Omega_{c}C^{\dag}_{j}C_{j}
$
and now
$
|\phi_{j}\rangle=e_{c}C^{\dag}_{j}|\emptyset\rangle
$.
Because of loss, the system would be nonconservative and corresponding operators
would be non-Hermitian. The communication relation of $C_{j}$ reads
$[C_{j},C^{\dag}_{j'}]=(1+i\frac{\gamma_{c}}{\omega_{c}})\delta_{jj'}$.
Recognizing $\frac{\gamma_{c}}{\omega_{c}}$ is in order of $\frac{1}{Q}$,
with $Q$ being the quality factor of individual cavity.
The bosonic communication relation is therefore approximately satisfied for
the high-$Q$ cavity, which is meeting in most experiments about
cavity quantum electrodynamics(QED).

The complex eigenfrequency underlines the facts that, on one hand,
dissipation is the inherent property for realistic cavity.
When a photon with certain frequency has been injected into a dissipative cavity,
the composite system of cavity-filed plus environment cannot be characterized merely
by the frequency of the injected photon, however, we must take the impacts of environment
into account. On the other hand, in general we are not concerned the
time evolution of bath. In this way, the array of dissipative cavities
can be regarded as a configuration consisted of quasi-bosons.
Quite similar operations can be performed on atom to introduce another
kind of quasi-boson described by $A_{j}$ with the frequency
$\Omega_{a}=\omega_{a}-i\gamma_{a}$,
where $\gamma_{a}$ is the atomic decay rate.

We can therefore rephrase Hamiltonian (1) with the
renormalized terms,
\begin{equation}
H=\sum_{j}H^{eff}_{j}-\sum_{<j,j'>}\kappa_{jj'}C^{\dag}_{j}C_{j'}-\sum_{j}\mu n_{j},
\end{equation}
with now
$H^{eff}_{j}=\Omega_{a}A^{\dag}_{j}A_{j}+\Omega_{c}C^{\dag}_{j}C_{j}+\beta(A^{\dag}_{j}C_{j}+h.c.)$
and
$n_{j}=A^{\dag}_{j}A_{j}+C^{\dag}_{j}C_{j}$.
One key feature of Hamiltonian (8) is now the losses
describe by leaky rates $\gamma_{a}$ and $\gamma_{c}$
but not by field oscillations. Without having to mention the external degrees of freedom,
this effective treatment would be of great conceptual and, moreover,
computational advantage rather than the general treatment as Hamiltonian (1).
A more microcosmic consideration points out that,
in cavity QED region, since the atom is dressed by cavity field,
the atom and field act as a whole subject to a total decay rate $\Gamma$~[\onlinecite{decay}].
In particular,
$\Gamma=n(\gamma_{a}+\gamma_{c})$
for
$\Delta=0$.

To gain insight over the role of dissipation in the SF-Mott phase transition, we use
a mean field approximation which could give reliable results if the system
is at least two-dimensional~[\onlinecite{MF}].
We introduce a superfluid parameter,
$\psi=Re\langle B_{j}\rangle=Re\langle B^{\dag}_{j}\rangle$.
In the present case, the expected value of $B_{j}(B^{\dag}_{j})$
is in general complex with the formation
$
B_{j}=\psi-i\psi_{\gamma}
(B^{\dag}_{j}=\psi+i\psi_{\gamma})
$.
$\psi_{\gamma}$ is a solvable small quantity as a function of
$\gamma_{a}$ and $\gamma_{c}$, and vanishes in the limit of no loss.
Using the decoupling approximation,
$
B^{\dag}_{j}B_{j'}=\langle B^{\dag}_{j}\rangle B_{j'}
+\langle B_{j'}\rangle B^{\dag}_{j}
-\langle B^{\dag}_{j}\rangle\langle B^{\dag}_{j}\rangle
$,
the resulting mean-field Hamiltonian can be written as a sum over single sites,
\begin{equation}
H^{MF}=\sum_{j}
\{
H^{eff}_{j}-z\kappa\psi(B^{\dag}_{j}+B_{j})+z\kappa|\psi|^{2}-\mu n_{j}+O(\psi^{2}_{\lambda})
\},
\end{equation}
where we have set the intercavity hopping rate
$\kappa_{jj'}=\kappa$ for all nearest-neighbors with $z$ labelling
the number. Corresponding to the order parameter in related closed
system, whether $\psi$ vanishes or has a finite value can be used
to identify the Mott-like and the SF-like regions.

$\psi$ can be examined analytically in terms of second-order perturbation theory,
with respect to the dressed basis.
For energetically favorable we assume each site is prepared in
the negative branch of dressed state.
But because the dressed basis is defined on $n\geq1$,
a ground state $|0\rangle$ with the energy $E_{|0\rangle}=0$ need to be supplemented,
\begin{equation}
\psi=e^{-\Gamma t}\sqrt{-\frac{\chi}{z\kappa\Theta}}.
\end{equation}
$\chi$ and $\Theta$ are functions of all the parameters of the whole system.
Since the evanescent parameter $\kappa$ is a typical small quantity
in systems of coupled cavities,
the perturbation theory gives good qualitative and quantitative descriptions
comparing to the numerically results
given by explicitly diagonalizing~[\onlinecite{SC2,SF-MI}].

Arguably the most interesting situation is the photon-photon interactions are
maximized, namely, cavities on resonant with atoms and
with one initial excitations per site~[\onlinecite{JCH}]. In addition,
$
\Gamma=\gamma_{a}+\gamma_{c}=\gamma
$.
With
$
F_{1}=\omega_{c}-\beta-\mu
$
and
$
F_{2}=-\omega_{c}+(\sqrt{2}-1)\beta+\mu
$,
in eq. (10),
$
\Theta=\frac{1}{2F^{2}_{1}+2\gamma^{2}}
+\frac{3+2\sqrt{2}}{4F^{2}_{2}+4\gamma^{2}}>0
$,
and
\begin{equation}
\chi=\frac{F_{1}}{2F^{2}_{1}+2\gamma^{2}}
+\frac{(3+2\sqrt{2})F_{2}}{4F^{2}_{2}+4\gamma^{2}}
+\frac{1}{z\kappa e^{-2\gamma t}}.
\end{equation}
In the absence of loss, one can recognize $\chi=0$ is the well known
self-consistent equation and therefore distinguish the SF phase and Mott phase.
Nevertheless, the coupling to environment inducing a non-equilibrium dynamics,
and there is thus no strict phase exist.
However, there remains a time-dependent boundary given by $\psi$
which can distinguish two slowly decayed and qualitatively different regions,
i.e. the SF-like and Mott-like regions.

In what follows, we discuss the physics of the transition between
these regions from two aspects. First, we start in the Mott phase
and discuss the impacts of dissipation on transition ratio
$(\frac{z\kappa}{\beta})_{c}$. Consider the initial state is deep
in the Mott phase, $ \frac{z\kappa}{\beta}=0 $, and we continually
increase the intercavity coupled rate. For the related ideal case,
we will reach the SF phase at $
\frac{z\kappa}{\beta}=(\frac{z\kappa}{\beta})'_{c}\simeq0.16 $.
However, in the presence of dissipation, owing to the continuous
leakage of coherence the effective tunnelling energy will be lower
than what we set. Stated in another way, the photon hopping rate
not only dependent on the coupling parameter between cavities, but
also on the coherence inside cavities. As shown in Fig. 2, one
cannot expect the appearance of photon hopping at
$\frac{z\kappa}{\beta}\simeq0.16$. We must continuously increase
$\kappa$ and the rate of increase must faster than the decay of
coherence. Only in such way can the transition occur before the
system reach the steady state. However, even we start exactly at $
\frac{z\kappa}{\beta}=(\frac{z\kappa}{\beta})'_{c} $, we also
cannot expect we are in the SF state. As the dissipation is the
inherent nature of open quantum system, the critical ration $
(\frac{z\kappa}{\beta})_{c} $, even at $t=0$, is modified by an
increment $ \simeq\frac{2\gamma^{2}}{\beta^{2}} $. This can be
understood in a simplified picture. If there was no excitation in
system, i.e. we are in $|\emptyset\rangle$, the decay would not
happen. However, the open quantum system is inherently different
from the correspondingly perfected system.
\begin{figure}
\includegraphics[clip]{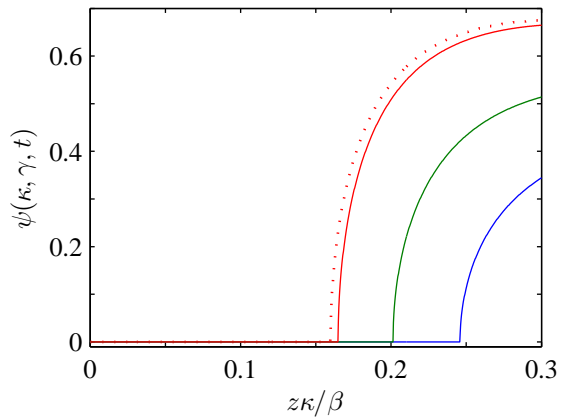}
\caption{
\label{fig:epsart}
The transition from Mott-like to SF-like region.
Influences of dissipation on the critical ratio
depend on the leaky rate $\gamma$
(the dot and solid line for $\frac{\gamma}{\beta}=0$, $0.05$, respectively)
and the time one reaches the transition
(the red, green, and blue lines for $t=0$, $0.1\gamma^{-1}$,
$0.2\gamma^{-1}$, respectively).
}
\end{figure}

In contrast, we start with the SF phase in the vicinity of
transition ratio and track the time evolution of system. The
exponential term in Eq.(10), $e^{-\Gamma t}$, indicates the
expected decay of $\psi$ and related physics quantities. However,
the more important impacts of the presence of external environment
are revealed by $\chi$. As illustrated in Fig. 3, for
$t\ll\beta^{-1}$, $\psi$ has a slightly decrease scaled by
$\frac{\gamma^{2}}{\beta^{2}}$. For $t>\beta^{-1}$, the leader
term is $z\kappa e^{-2\gamma t}$. It is this term pronounces the
decrease of effective tunnelling energy which, mathematically,
originates from the non-diagonal element in Hamiltonian (9)
respecting to the occupation number basis. Consequently, a photon
hopping rate $ \frac{z\kappa}{\beta} $ given initially in SF phase
will cross the critical point at a time $
t_{c}\simeq-\frac{1}{2\gamma}\ln\frac{\kappa}{\kappa'_{c}} $.
Before $t_{c}$, a non-local region is still recognized as
non-local. The dissipation does not change the fundamental nature
of the system, albeit with the decrease of long-range coherence.
Nevertheless, beyond $t_{c}$, the superfluidity breaks down and
gives rise to a localized wavefunction, i.e. a transition to
Mott-like region do occur. An analogous effect is recently
described in optical lattice system, where a time-dependent
tunneling rate controlled by external filed leads to a sweep from
the SF to the Mott phase~[\onlinecite{sweep}]. Differently, in our
case the decay of long-range order is resulted from the leakage of
coherence, and the total number of excitations is not
conservative. As a result, there remains a statistical fluctuation
acted on each site.
\begin{figure}
\includegraphics[clip]{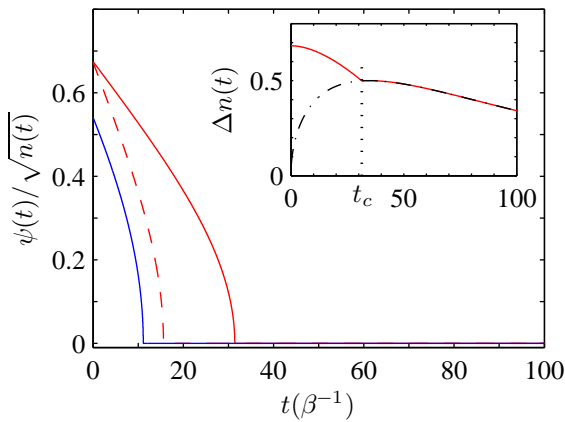}
\caption{
\label{fig:epsart}
The temporal decrease of the superfluidity
and the photon number fluctuation on each site for a certain
initial state(inset).
For a initial SF state
(the red line for $\frac{z\kappa}{\beta}=0.3$
and blue line for $\frac{z\kappa}{\beta}=0.2$),
before $t_{c}$ the long-range order decays continuously
and the fluctuation on each site is a total effect of photon hopping
and photon leakage
(the solid line for $\frac{\gamma}{\beta}=0.01$
and dash line for $\frac{\gamma}{\beta}=0.02$).
Beyond $t_{c}$, the superfluidity breaks down and the related photon
number fluctuation behaves as the fluctuation of a Mott-like state
(dot-dash line).
}
\end{figure}

In conclusion, we have shown analytically the features of
SF-Mott phase transition in the coupled cavity arrays
in the presence of dissipation. Our analysis fully takes
into account the intrinsically dissipative nature of
open quantum many-body system, and identifies how dissipation
and decoherence would come into play.
For further experiment realization, we predict there is
a localized tendency characterized by the suppression to
the superfluidity.

This work was supported by NSFC under grants
Nos. 10704031, 10874235, 10934010 and 60978019,
the NKBRSFC under grants
Nos. 2009CB930701, 2010CB922904 and 2011CB921500,
and the Fundamental Research Funds for the Central Universities under grant
No. lzujbky-2010-75.

\end{document}